# Solution of the spin and pseudo-spin symmetric Dirac equation in 1+1 space-time using the tridiagonal representation approach

I. A. Assi[a], A. D. Alhaidari[b], and H. Bahlouli[a,b]

[a] *Physics Department, King Fahd University of Petroleum & Minerals, Dhahran 31261, Saudi Arabia*
[b] *Saudi Center for Theoretical Physics, P.O. Box 32741, Jeddah 21438, Saudi Arabia*

**Abstract**: The aim of this work is to find exact solutions of the Dirac equation in 1+1 space-time beyond the already known class. We consider exact spin (and pseudo-spin) symmetric Dirac equations where the scalar potential is equal to plus (and minus) the vector potential. We also include pseudo-scalar potentials in the interaction. The spinor wavefunction is written as a bounded sum in a complete set of square integrable basis, which is chosen such that the matrix representation of the Dirac wave operator is tridiagonal and symmetric. This makes the matrix wave equation a symmetric three-term recursion relation for the expansion coefficients of the wavefunction. We solve the recursion relation exactly in terms of orthogonal polynomials and obtain the state functions and corresponding relativistic energy spectrum and phase shift.

We are honored to dedicate this work to Prof. Hashim A. Yamani on the occasion of his 70[th] birthday.

## 1. Introduction

The Dirac wave equation is used to describe the dynamics of spin one-half particles at high energies (but below the threshold of pair creation) in relativistic quantum mechanics. It is a relativistically covariant linear first order differential equation in space and time for a multi-component spinor wavefunction. This equation is consistent with both the principles of quantum mechanics and the theory of special relativity [1]. The physics and mathematics of the Dirac equation are very rich, illuminating and gave birth to the theoretical foundation for different physical phenomena that were not observed in the non-relativistic regime. Among others, we can cite the prediction of electron spin, the existence of antiparticles and tunneling through very high barriers, the so-called Klein tunneling [2]. In addition, Dirac equation appears at a lower energy scale in graphene (2 D array of carbon atoms), wherein the behavior of electrons is modeled by 2 D massless Dirac equation, the so-called Dirac-Weyl equation [3]. However, despite its fundamental importance in physics, exact solutions of the Dirac equation were obtained only for a very limited class of potentials [4].

In this paper, we study situations with spin or pseudo-spin symmetry which are SU(2) symmetries of the Dirac equation that have different applications especially in nuclear physics [5]. The spin symmetric case is generally defined for situations where $\Delta(r) = S(r) - V(r) = C_s$, where $C_s$ is a real constant while $S(r)$ and $V(r)$ are the scalar and vector components of the potential, respectively. Spin symmetry has been used to explain the suppression of spin-orbit splitting of meson states with heavy and light quarks. Pseudospin symmetry occurs when $\Sigma(r) = S(r) + V(r) = C_p$, where $C_p$ is a real constant parameter. This latter symmetry was used to explain the near degeneracy of some single particle levels near the Fermi surface. Here, we will restrict our study to the exact symmetry where $C_s = C_p = 0$, that is when $S(r) = \pm V(r)$. Aside from their physical applications, these symmetries allow the decoupling of the upper and lower spinor components of the Dirac equation transforming it into a Schrödinger-like equation for each of the two components. This makes it mathematically easier to obtain analytic solutions of the original wave equation for certain potential configurations. In addition to the scalar and vector potentials, we also include a pseudo-scalar component to the potential configuration.

Exact solutions of the Dirac equation are of great benefit both from the theoretical and applied point of view. Analytic solutions allow for a better understanding of physical phenomena and establish the necessary correspondence between relativistic effects and their non-relativistic analogues. In this spirit, we would like to revisit the one-dimensional Dirac equation and investigate all potentially solvable class of interactions using the tridiagonal representation approach (TRA) [6]. The hope is to be able to enlarge the conventional class of solvable potentials of the Dirac equation. The basic idea of this approach is to write the spinor wave function as a bounded infinite series with respect to a suitably chosen square integrable basis function. That is, $|\psi_\varepsilon(x)\rangle = \sum_m f_m(\varepsilon)|\phi_m(x)\rangle$, where $\{f_m(\varepsilon)\}_{m=0}^{\infty}$ is a set of expansion coefficients that are functions of the energy $\varepsilon$ and potential parameters whereas $\{|\phi_m(x)\rangle\}_{m=0}^{\infty}$ is a complete set of spinor basis functions that carry only kinematic information. Using this form of the spinor wavefunction, the stationary wave equation becomes $(H - \varepsilon)|\psi\rangle = J|\psi\rangle = 0$, where $H$ is the Dirac Hamiltonian. We require that the matrix representation of the wave operator, $J_{n,m} = \langle\phi_n|(H-\varepsilon)|\phi_m\rangle$, be tridiagonal and symmetric so that the action of the wave operator on the elements of the basis is allowed to take the general form $(H-E)|\phi_n\rangle \sim |\phi_n\rangle + |\phi_{n-1}\rangle + |\phi_{n+1}\rangle$. To achieve this requirement, we were obliged to use the kinetic balance equation that relates the upper and the lower spinor components transforming the wave equation into the following three-term recursion relation for $\{f_m(\varepsilon)\}_{m=0}^{\infty}$ [7,8]:

$$J_{n,n} f_n(\varepsilon) + J_{n,n-1} f_{n-1}(\varepsilon) + J_{n,n+1} f_{n+1}(\varepsilon) = 0, \qquad (1.1)$$

Thus, the problem now is reduced to solving this three-term recursion relation which is equivalent to solving the original problem since $\{f_m(\varepsilon)\}_{m=0}^{\infty}$ contain all physical information (both structural and dynamical) about the system. Of course, this equation can be solved in different ways in mathematics [7,8]. For example, Eq. (1.1) could be written in a form that allows for direct comparison to well-known orthogonal polynomials. However, in other situations this recursion relation does not correspond to any of the known orthogonal polynomials hence giving rise to new class of orthogonal polynomials. The remaining challenge will then be to extract from these recursion relations the important properties of the associated orthogonal polynomials such as the weight function, generating function, spectrum formula, asymptotics, zeroes, etc. [9, 10].

The organization of this work goes as follows. In section 2, we present our mathematical formulation of the problem for the spin and pseudo-spin symmetric situations. Then, in sections 3 and 4 we present different examples of solvable potentials. Lastly, we conclude our work in the 5$^{th}$ section.

## 2. Formulation

In the relativistic units, $\hbar = c = 1$, the most general linear massive Dirac equation in 1+1 space-time dimension with time-independent potentials reads as follows [1]:

$$\{\gamma^\mu[i\partial_\mu - A_\mu(x)] - S(x) - \gamma^5 W(x)\}\Psi(t,x) = M\Psi(t,x), \qquad (2.1)$$

where $\{\gamma^\mu\}_{\mu=0,1}$ are the two Dirac gamma matrices such that $(\gamma^0)^2 = 1$, $(\gamma^1)^2 = -1$ and $\gamma^0\gamma^1 = -\gamma^1\gamma^0$. $S$ is the scalar potential, $A_\mu = (V, U)$ is the (time, space) component of the two-vector potential, $W$ is the pseudo-scalar potential, and $\gamma^5 = i\gamma^0\gamma^1$. Choosing the minimum dimensional representation for the gamma matrices defined by $\gamma^0 = \begin{pmatrix} 1 & 0 \\ 0 & -1 \end{pmatrix}$ and $\gamma^1 = -i\begin{pmatrix} 0 & 1 \\ 1 & 0 \end{pmatrix}$, gives $\gamma^5 = \begin{pmatrix} 0 & 1 \\ -1 & 0 \end{pmatrix}$ and makes the Dirac equation (2.1) take the following matrix form

$$\begin{pmatrix} M + S(x) + V(x) & -\frac{d}{dx} + W(x) - iU(x) \\ \frac{d}{dx} + W(x) + iU(x) & -M - S(x) + V(x) \end{pmatrix} \begin{pmatrix} \psi^+(x) \\ \psi^-(x) \end{pmatrix} = \varepsilon \begin{pmatrix} \psi^+(x) \\ \psi^-(x) \end{pmatrix}, \qquad (2.2)$$

where $\varepsilon$ is the energy and we wrote the two-component spinor wavefunction as $\Psi(t,x) = e^{-i\varepsilon t}\begin{pmatrix} \psi^+(x) \\ \psi^-(x) \end{pmatrix} = e^{-i\varepsilon t}\psi(x)$. The space component of the vector potential, $U$, could be eliminated by the local gauge transformation $\psi(x) \to e^{-i\Lambda(x)}\psi(x)$ such that $d\Lambda/dx = U$. Therefore, from now on and for simplicity, we take $U = 0$. It should be noted that in 3+1 space-time with spherical symmetry, Eq. (2.2) with $W(x) \to W(x) + \frac{\kappa}{x}$ represents the radial Dirac equation with $x$ being the radial coordinate and the spin-orbit quantum number $\kappa = \pm 1, \pm 2, \pm 3, ...$ Now, the exact spin and pseudo-spin symmetric coupling correspond to $S = V$ and $S = -V$, respectively. We discuss below the positive energy solution of the spin symmetric coupling in the time-independent Dirac equation. The negative energy pseudo-spin symmetric solution follows from the spin symmetric one by a straightforward map, which will be derived below.

Now, for spin symmetric coupling the Dirac equation (2.2) with $U = 0$ reads as follows

$$\begin{pmatrix} 2V + M & -\frac{d}{dx} + W(x) \\ \frac{d}{dx} + W(x) & -M \end{pmatrix} \begin{pmatrix} \psi^+(x) \\ \psi^-(x) \end{pmatrix} = \varepsilon \begin{pmatrix} \psi^+(x) \\ \psi^-(x) \end{pmatrix}. \qquad (2.3)$$

Giving the following relation between the two spinor components for the positive energy solution space

$$\psi^-(x) = \frac{1}{\varepsilon + M}\left[\frac{d}{dx} + W(x)\right]\psi^+(x), \qquad (2.4)$$

where $\varepsilon \neq -M$. Substituting this expression of $\psi^-(x)$ in (2.3) gives the following Schrödinger-like second order differential equation for the upper spinor component

$$\left[-\frac{d^2}{dx^2} + W^2 - \frac{dW}{dx} + 2(\varepsilon + M)V + M^2 - \varepsilon^2\right]\psi^+(x) = 0. \qquad (2.5)$$

The objective now is to find a discrete square integrable spinor basis in which the matrix representation of the wave equation (2.3) becomes tridiagonal and symmetric so that the corresponding three-term recursion relation could be solved exactly for the expansion coefficients of the wavefunction and for as large a class of potentials as possible. As noted in the Introduction above, we write $\psi^\pm(x) = \sum_n f_n(\varepsilon)\phi_n^\pm(x)$, where $\{\phi_n^\pm(x)\}$ is a complete set of square integrable basis elements for the two wavefunction components and $\{f_n(\varepsilon)\}$ is an appropriate set of energy dependent functions. Now, in the TRA, we impose the requirement that the matrix representation of the Dirac wave operator $J_{nm} = \langle \phi_n | (H - \varepsilon) | \phi_m \rangle$ is tridiagonal and symmetric (with $\phi_n$ being a spinor whose components are $\phi_n^\pm$) so that the wave equation (2.3) becomes a three-term recursion relation for the expansion coefficients $\{f_n\}$:

$$J_{nm} = \langle \phi_n^+ | (2V + M - \varepsilon) | \phi_m^+ \rangle - (\varepsilon + M) \langle \phi_n^- | \phi_m^- \rangle \\ + \langle \phi_n^+ | \left( -\tfrac{d}{dx} + W \right) | \phi_m^- \rangle + \langle \phi_n^- | \left( \tfrac{d}{dx} + W \right) | \phi_m^+ \rangle \quad (2.6)$$

In line with the kinetic balance approach for the two spinor components, we relate the two components of the spinor basis using Eq. (2.4) as $\phi_n^- = \frac{1}{\varepsilon + M} \left( \frac{d}{dx} + W \right) \phi_n^+$. Using this and the fact that $\phi_n^+(x)$ vanishes at the boundaries of configuration space, then we can perform integration by parts and rewrite the above equation as

$$J_{nm} = 2 \langle \phi_n^+ | V | \phi_m^+ \rangle - (\varepsilon - M) \langle \phi_n^+ | \phi_m^+ \rangle + (\varepsilon + M)^{-1} \langle \phi_n^+ | \left[ -\frac{d^2}{dx^2} + W^2 - \frac{dW}{dx} \right] | \phi_m^+ \rangle. \quad (2.7)$$

Now, for the pseudo-spin symmetric coupling ($S = -V$), the Dirac equation (2.2) with $U = 0$ becomes

$$\begin{pmatrix} M & -\tfrac{d}{dx} + W(x) \\ \tfrac{d}{dx} + W(x) & 2V - M \end{pmatrix} \begin{pmatrix} \psi^+(x) \\ \psi^-(x) \end{pmatrix} = \varepsilon \begin{pmatrix} \psi^+(x) \\ \psi^-(x) \end{pmatrix}, \quad (2.8)$$

and the kinetic balance relation (2.4) is replaced by

$$\psi^+(x) = \frac{1}{M - \varepsilon} \left[ \frac{d}{dx} - W(x) \right] \psi^-(x), \quad (2.9)$$

where $\varepsilon \neq +M$. Substituting this in (2.8) gives the following Schrödinger-like second order differential equation for the lower spin component

$$\left[ -\frac{d^2}{dx^2} + W^2 + \frac{dW}{dx} + 2(\varepsilon - M)V + M^2 - \varepsilon^2 \right] \psi^-(x, \varepsilon) = 0, \quad (2.10)$$

Comparing equations (2.9) and (2.10) with the corresponding spin symmetric case, we obtain the following map

$$W \to -W, \; V \to -V, \; \varepsilon \to -\varepsilon, \text{ and } \psi^\pm \to \psi^\mp, \quad (2.11)$$

Applying this map to Eq. (2.3), we obtain

$$\begin{pmatrix} -2V + M & -\tfrac{d}{dx} - W(x) \\ \tfrac{d}{dx} - W(x) & -M \end{pmatrix} \begin{pmatrix} \psi^-(x) \\ \psi^+(x) \end{pmatrix} = -\varepsilon \begin{pmatrix} \psi^-(x) \\ \psi^+(x) \end{pmatrix}, \quad (2.12)$$

Multiplying this equation by $-1$ then from left by $\sigma_x = \begin{pmatrix} 0 & 1 \\ 1 & 0 \end{pmatrix}$, and noting that $\sigma_x^2 = 1$, we obtain

$$\begin{pmatrix} 0 & 1 \\ 1 & 0 \end{pmatrix} \begin{pmatrix} 2V - M & \tfrac{d}{dx} + W(x) \\ -\tfrac{d}{dx} + W(x) & M \end{pmatrix} \begin{pmatrix} 0 & 1 \\ 1 & 0 \end{pmatrix} \begin{pmatrix} 0 & 1 \\ 1 & 0 \end{pmatrix} \begin{pmatrix} \psi^- \\ \psi^+ \end{pmatrix} = \varepsilon \begin{pmatrix} 0 & 1 \\ 1 & 0 \end{pmatrix} \begin{pmatrix} \psi^- \\ \psi^+ \end{pmatrix}, \quad (2.13)$$

Giving and identical equation to the pseudo-spin symmetric Dirac equation (2.8). Thus, applying the map (2.11) on the positive energy spin symmetric solution gives the negative energy pseudo-spin symmetric solution.

The most general square-integrable basis takes the following form [6]:

$$\phi_m(x) = A_m w(y) P_m(y) \tag{2.14}$$

where $y = y(x)$, $A_m$ is a normalization constant and $P_m(y)$ is a polynomial of a degree $m$ in $y$. Whereas, $w(y)$ is a positive function that vanishes on the boundaries of the original configuration space with coordinate $x$ and has a form similar to the weight function associated with the polynomial. In our present work, we will be using two sets of bases:

1. The Laguerre basis, where $P_m(y) = L_m^\nu(y)$ are the Laguerre polynomials with $y \geq 0$ and $w(y) = y^\alpha e^{-\beta y}$.
2. The Jacobi basis where $P_m(y) = P_m^{(\mu,\nu)}(y)$ are the Jacobi polynomials with $y \in [-1, +1]$ and $w(y) = (1-y)^\alpha (1+y)^\beta$.

In the following two sections, we obtain the exact positive energy solution of the spin symmetric Dirac equation (2.3) in the Laguerre and Jacobi bases by giving the expansion coefficients $\{f_n\}$ in terms of orthogonal polynomials in the energy variable. The asymptotics of these polynomials give the phase shift of the continuous energy scattering states and the spectrum of the discrete energy bound states.

## 3. Solution in the Laguerre basis

Let $y(x)$ be a transformation from the real configuration space with coordinate $x$ to a new dimensionless coordinate $y$ such that $y \geq 0$. A complete set of square integrable functions as basis for the wavefunction in the new $y$-space that also satisfy the desirable boundary conditions (vanish at the boundaries) could be chosen as follows

$$\phi_n^+(x) = A_n y^\alpha e^{-\beta y} L_n^\nu(y), \tag{3.1}$$

where $A_n = \sqrt{\Gamma(n+1)/\Gamma(n+\nu+1)}$ and the real parameters are such that $\beta > 0$ and $\nu > -1$. Substituting (3.1) into Eq. (2.4), as applied to the basis set, and using the differentiation chain rule $\frac{d}{dx} = y' \frac{d}{dy}$, we obtain

$$\phi_n^-(x) = \frac{A_n}{\varepsilon + M} y' y^{\alpha-1} e^{-\beta y} \left[ y \frac{d}{dy} + \alpha - \beta y + \frac{y}{y'} W(y) \right] L_n^\nu(y), \tag{3.2}$$

where the prime stands for the derivative with respect to $x$. It is required that in function space $\phi_n^-(x)$ must be nearest neighbor to $\phi_n^+(x)$. That is, the tridiagonal requirement on equation (3.2) means that $\phi_n^-(x)$ should be expressed as a linear combination of terms in $\phi_n^+(x)$ and $\phi_{n\pm1}^+(x)$. The differential property of the Laguerre polynomial, $y \frac{d}{dy} L_n^\nu = n L_n^\nu - (n+\nu) L_{n-1}^\nu$ and its recursion relation, $y L_n^\nu = (2n+\nu+1) L_n^\nu - (n+\nu) L_{n-1}^\nu - (n+1) L_{n+1}^\nu$ show that this could be achieved if we impose the constraint that $\frac{y}{y'} W(y)$ be a linear function in $y$. That is, $\frac{y}{y'} W(y) = \rho y + \sigma$ where $\rho$ and $\sigma$ are dimensionless potential parameters. Thus, we can rewrite Eq. (3.2) as follows

$$\phi_n^-(x) = \frac{y'y^{-1}}{\varepsilon + M}\left\{\left[\left(\rho - \beta + \tfrac{1}{2}\right)(2n + v + 1) + \alpha + \sigma - \tfrac{v+1}{2}\right]\phi_n^+(x)\right.$$
$$\left. -(\rho - \beta + 1)\sqrt{n(n+v)}\phi_{n-1}^+(x) - (\rho - \beta)\sqrt{(n+1)(n+v)}\phi_{n+1}^+(x)\right\} \qquad (3.3)$$

In the following two subsections, we consider the two physical scenarios found in Appendix A that correspond to equations (A6a) and (A6b), respectively.

## 3.1 The $q = 0$ scenario of (A6a):

In this scenario, the vector potential is $V(y) = y^{2a-1}(A + By)$ and the pseudo-scaler potential is $W(y) = \lambda y^{a-1}(\rho y + \sigma)$, where $a$, $A$ and $B$ are real parameters introduced in the Appendix. Moreover, the basis parameters are $v^2 = (2\sigma + 1 - a)^2$ and $2\alpha = v + 1 - a$. There are two configurations in this scenario: one corresponds to $a = 0$ where $y(x) = \lambda x$ and the other corresponds to $a = \tfrac{1}{2}$ where $y(x) = (\lambda x/2)^2$.

For the first configuration, we can write

$$V(r) = V_0 + \frac{Z}{r}, \text{ and } W(r) = W_0 + \frac{\kappa}{r}, \qquad (3.1.1)$$

where we have replaced $x$ by the radial coordinate $r$ and $Z$ stands for the electric charge whereas $\kappa$ is the spin-orbit coupling which assumes the values $\pm 1, \pm 2, \pm 3, \ldots$. Therefore, the parameters in the Dirac wave operator matrix (A7a) are as follows:

$$\rho = W_0/\lambda, \ \sigma = \kappa, \ A = \lambda Z, \ B = V_0 \text{ and } v^2 = (2\kappa + 1)^2. \qquad (3.1.2)$$

These parameter assignments are physically motivated and will be supported by the results obtained below. Additionally, to make the vector potential pure Coulombic and vanish at infinity we can freely choose $V_0 = 0$. Substituting these parameters in (A7a), then the Dirac equation $H|\psi\rangle = \varepsilon|\psi\rangle$ results in the following symmetric three-term recursion relation for the expansion coefficients of the wave function

$$\left[\frac{2Z}{\lambda}(\varepsilon + M) + \frac{2W_0}{\lambda}\kappa\right]f_n = -(2n + v + 1)\left[\frac{\varepsilon^2 - \tilde{M}^2}{\lambda^2} - \frac{1}{4}\right]f_n$$
$$+ \left[\frac{\varepsilon^2 - \tilde{M}^2}{\lambda^2} + \frac{1}{4}\right]\left[\sqrt{n(n+v)}f_{n-1} + \sqrt{(n+1)(n+v+1)}f_{n+1}\right] \qquad (3.1.3)$$

where $\tilde{M}^2 = M^2 + W_0^2$. Dividing by $\frac{\varepsilon^2 - \tilde{M}^2}{\lambda^2} + \frac{1}{4}$ and writing $f_n(\varepsilon) = f_0(\varepsilon)P_n(\varepsilon)$, makes this a recursion relation for $P_n(\varepsilon)$ with $P_0(\varepsilon) = 1$. We compare it to that of the orthonormal version of the two-parameter Meixner-Pollaczek polynomial $P_n^\mu(z, \theta)$ that reads

$$(z \sin \theta) P_n^\mu(z, \theta) = -[(n + \mu) \cos \theta] P_n^\mu(z, \theta)$$
$$+ \tfrac{1}{2}\sqrt{n(n + 2\mu - 1)} P_{n-1}^\mu(z, \theta) + \tfrac{1}{2}\sqrt{(n+1)(n+2\mu)} P_{n+1}^\mu(z, \theta) \qquad (3.1.4)$$

where the Meixner-Pollaczek polynomial is defined as

$$P_n^\mu(z, \theta) = \sqrt{\frac{\Gamma(n+2\mu)}{\Gamma(2\mu)\Gamma(n+1)}}\, e^{in\theta}\, {}_2F_1\left(\genfrac{}{}{0pt}{}{-n, \mu + iz}{2\mu}\bigg|1 - e^{-2i\theta}\right), \qquad (3.1.5)$$

with $z \in [-\infty, +\infty]$, $\mu > 0$ and $0 < \theta < \pi$. Thus, we conclude that $P_n(\varepsilon) = P_n^\mu(z, \theta)$ where

$$\mu = \tfrac{1}{2}(\nu+1) = \begin{cases} \kappa+1 &, \kappa>0 \\ -\kappa &, \kappa<0 \end{cases}, \quad \cos\theta = \frac{\varepsilon^2 - \tilde{M}^2 - \tfrac{1}{4}\lambda^2}{\varepsilon^2 - \tilde{M}^2 + \tfrac{1}{4}\lambda^2}, \tag{3.1.6}$$

and $z = \dfrac{Z(\varepsilon+M)+\kappa W_0}{\sqrt{\varepsilon^2-\tilde{M}^2}}$ requiring that $\varepsilon^2 \geq \tilde{M}^2$. The phase shift is obtained by using the asymptotics ($n \to \infty$) formula of the Meixner-Pollaczek polynomial. It reads $\delta^\mu(z) = \arg\Gamma(\mu+iz) + \mu(\theta - \pi/2)$ giving

$$\delta(\varepsilon) = \arg\Gamma\left[\frac{\nu+1}{2} + i\frac{Z(\varepsilon+M)+\kappa W_0}{\sqrt{\varepsilon^2-\tilde{M}^2}}\right] + \frac{\nu+1}{2}\left[\theta(\varepsilon) - \frac{\pi}{2}\right], \tag{3.1.7}$$

The continuous energy wavefunction becomes $\psi^\pm(r) = f_0(\varepsilon)\sum_n P_n^\mu(z,\theta)\phi_n^\pm(r)$. On the other hand, the spectrum formula of the Meixner-Pollaczek polynomial that reads $z_n^2 = -(n+\mu)^2$ gives the following relativistic energy spectrum

$$\varepsilon_n^2 = \tilde{M}^2 - \left[\frac{Z(\varepsilon_n+M)+\kappa W_0}{n + \frac{\nu+1}{2}}\right]^2, \tag{3.1.8}$$

which is a quadratic equation that could easily be solved for $\varepsilon_n$. With $W_0 = 0$, it is identical to the energy spectrum of the spin-symmetric Dirac-Coulomb problem (i.e., with equal scalar and vector Coulomb potentials). The $m^{\text{th}}$ bound state energy wavefunction, will be written in terms of the Meixner polynomial $M_n^\mu(m;\gamma)$, which is the discrete version of the Meixner-Pollaczek polynomial, as $\psi^\pm(r) = f_0(\varepsilon_m)\sum_n M_n^\mu(m;\gamma)\phi_n^\pm(r)$. The orthonormal version of this polynomial is defined as

$$M_n^\mu(m;\gamma) = \sqrt{\frac{\Gamma(n+2\mu)\gamma^n}{\Gamma(2\mu)\Gamma(n+1)}}\,_2F_1\!\left(\begin{array}{c}-n,-m\\2\mu\end{array}\bigg|1-\gamma^{-1}\right), \tag{3.1.9}$$

where $0 < \gamma < 1$ and $\mu > 0$. It satisfies the following recursion relation

$$(1-\gamma)m M_n^\mu(m;\gamma) = [n(1+\gamma) + 2\mu\gamma] M_n^\mu(m;\gamma) \\ - \sqrt{n(n+2\mu-1)\gamma}\, M_{n-1}^\mu(m;\gamma) - \sqrt{(n+1)(n+2\mu)\gamma}\, M_{n+1}^\mu(m;\gamma) \tag{3.1.10}$$

where $\gamma^{-1} = \cosh\theta$.

For the second configuration, $y(x) = (\lambda x/2)^2$ and we can write

$$V(r) = V_0 + \frac{1}{4\lambda}\Omega^4 r^2, \text{ and } W(r) = \frac{\kappa}{r} + \frac{1}{2}\omega^2 r, \tag{3.1.11}$$

where we have also replaced $x$ by the radial coordinate $r$ and $\kappa$ is the spin-orbit coupling constant. $\Omega$ is the vector oscillator frequency whereas the pseudo-scalar oscillator frequency is $\omega$. These parameter assignments are motivated by physical expectations as will be justified by the results obtained below. Without loss of generality, we can always choose $V_0 = 0$. Therefore, the parameters in the Dirac wave operator matrix (A7a) are as follows:

$$\rho = (\omega/\lambda)^2, \quad \sigma = \tfrac{1}{2}\kappa, \quad A = 0, \quad B = \Omega^4/\lambda^3 \text{ and } \nu^2 = (\kappa + \tfrac{1}{2})^2. \tag{3.1.12}$$

Substituting these parameters in (A7a), results in the following symmetric three-term recursion relation for the expansion coefficients of the wave function

$$\left[\frac{\varepsilon^2 - M^2}{\lambda^2} - \frac{\omega^2}{\lambda^2}\left(\kappa - \tfrac{1}{2}\right)\right] f_n = (2n + \nu + 1)\left[2\frac{\Omega^4}{\lambda^4}\frac{\varepsilon + M}{\lambda} + \frac{\omega^4}{\lambda^4} + \frac{1}{4}\right] f_n$$
$$+ \left[2\frac{\Omega^4}{\lambda^4}\frac{\varepsilon + M}{\lambda} + \frac{\omega^4}{\lambda^4} - \frac{1}{4}\right]\left[\sqrt{n(n+\nu)}\, f_{n-1} + \sqrt{(n+1)(n+\nu+1)}\, f_{n+1}\right]$$
(3.1.13)

Dividing by $\frac{1}{4} - 2\frac{\Omega^4}{\lambda^4}\frac{\varepsilon + M}{\lambda} - \frac{\omega^4}{\lambda^4}$ and writing $f_n(\varepsilon) = f_0(\varepsilon) P_n(\varepsilon)$, makes this a recursion relation for $P_n(\varepsilon)$ with $P_0(\varepsilon) = 1$. Comparing with that of the Meixner-Pollaczek polynomial (3.1.4) dictates that $P_n(\varepsilon) = P_n^\mu(z, \theta)$ with

$$\mu = \tfrac{1}{2}(\nu + 1) = \tfrac{1}{2}\begin{cases} \kappa + \tfrac{3}{2}, & \kappa = -1, +1, +2, \dots \\ -\kappa + \tfrac{1}{2}, & \kappa = -1, -2, -3, \dots \end{cases}, \quad \cos\theta = \frac{\tfrac{1}{4}\lambda^4 + 2\Omega^4 \lambda^{-1}(\varepsilon + M) + \omega^4}{\tfrac{1}{4}\lambda^4 - 2\Omega^4 \lambda^{-1}(\varepsilon + M) - \omega^4}, \quad (3.1.14)$$

and $z = \dfrac{\omega^2\left(\kappa - \tfrac{1}{2}\right) - \left(\varepsilon^2 - M^2\right)}{2\sqrt{-2\Omega^4 \lambda^{-1}(\varepsilon + M) - \omega^4}}$. It is obvious that for positive energy where $\varepsilon \geq M$, these assignments violate reality since $z$ becomes pure imaginary and $|\cos\theta| > 1$. Thus, we are forced to make the replacement $z \to iz$ and $\theta \to i\theta$ changing the trigonometric functions in (3.1.14) to hyperbolic and making the asymptotic wavefunction vanish since to the oscillation factor $e^{in\theta}$ in (3.1.5) changes into a decaying factor $e^{-n\theta}$. All of this imply that there are no continuous energy scattering states but only discrete energy bound states. This, of course, is an expected result for the isotropic oscillator whose energy spectrum is confirmed by using the spectrum formula of the Meixner-Pollaczek polynomial that gives

$$\left[\varepsilon_n^2 - M^2 - \omega^2\left(\kappa - \tfrac{1}{2}\right)\right]^2 = (2n + \nu + 1)^2 \left[2\Omega^4 \lambda^{-1}(\varepsilon_n + M) + \omega^4\right]. \quad (3.1.15)$$

With $\Omega = 0$, this is identical to the energy spectrum of the Dirac-oscillator problem. The corresponding $m^{\text{th}}$ bound state wavefunction will be written in terms of the discrete version of the Meixner-Pollaczek polynomial as $\psi_m^\pm(r) = f_0(\varepsilon_m) \sum_n M_n^\mu(m; \gamma) \phi_n^\pm(r)$.

### 3.2 The $q = 1$ scenario of (A6b):

In this scenario, the vector potential is $V(y) = y^{2a-2}(A + By)$ and the pseudo-scaler potential is $W(y) = \lambda y^{a-1}(\rho y + \sigma)$ such that $\rho^2 = \tfrac{1}{4}$. Moreover, the basis parameter $\nu$ is to be determined later by physical constraints whereas $2\alpha = \nu + 2 - a$. There are two configurations in this scenario: one corresponds to $a = 1$ where $y(x) = e^{\lambda x}$ with $-\infty < x < +\infty$ and the other corresponds to $a = \tfrac{1}{2}$ where $y(x) = (\lambda x/2)^2$.

For the first configuration, we can write
$$V(x) = V_0 + V_1 e^{\lambda x}, \text{ and } W(x) = W_0 + W_1 e^{\lambda x}, \quad (3.2.1)$$
To make the vector potential vanish at infinity we can freely choose $V_0 = 0$. Therefore, the parameters in the Dirac wave operator matrix (A7b) are as follows:
$$\rho = W_1/\lambda, \; \sigma = W_0/\lambda, \; A = 0, \; B = V_1 \text{ and } W_1^2 = (\lambda/2)^2, \quad (3.2.2)$$
Substituting these parameters in (A7b), results in the following symmetric three-term recursion relation for the expansion coefficients of the wave function

$$\frac{\varepsilon^2 - M^2 - W_0^2}{\lambda^2} f_n = \left\{ (2n+\nu+1)\left[ n + \frac{\nu}{2} + 1 + \frac{2V_1}{\lambda^2}(\varepsilon+M) + \frac{W_1}{\lambda}\left(\frac{2W_0}{\lambda} - 1\right) \right] - n - \frac{1}{4}(\nu+1)^2 \right\} f_n$$
$$- \left[ n + \frac{\nu}{2} + 1 + \frac{2V_1}{\lambda^2}(\varepsilon+M) + \frac{W_1}{\lambda}\left(\frac{2W_0}{\lambda} - 1\right) \right]\sqrt{(n+1)(n+\nu+1)} f_{n+1} - \left[ n + \frac{\nu}{2} + \frac{2V_1}{\lambda^2}(\varepsilon+M) + \frac{W_1}{\lambda}\left(\frac{2W_0}{\lambda} - 1\right) \right]\sqrt{n(n+\nu)} f_{n-1}$$
(3.2.3)

Writing $f_n(\varepsilon) = f_0(\varepsilon) P_n(\varepsilon)$, makes this a recursion relation for $P_n(\varepsilon)$ with $P_0(\varepsilon) = 1$. We compare it to that of the orthonormal version of the three-parameter continuous dual Hahn polynomial $S_n^\xi(z^2;\gamma,\tau)$ that reads

$$z^2 S_n^\xi = \left[ (n+\xi+\gamma)(n+\xi+\tau) + n(n+\gamma+\tau-1) - \xi^2 \right] S_n^\xi$$
$$- \sqrt{n(n+\gamma+\tau-1)(n+\xi+\gamma-1)(n+\xi+\tau-1)}\, S_{n-1}^\xi \quad (3.2.4)$$
$$- \sqrt{(n+1)(n+\gamma+\tau)(n+\xi+\gamma)(n+\xi+\tau)}\, S_{n+1}^\xi$$

where the continuous dual Hahn polynomial is defined as

$$S_n^\xi(z^2;\gamma,\tau) = \sqrt{\frac{(\xi+\gamma)_n(\xi+\tau)_n}{n!(\gamma+\tau)_n}}\, {}_3F_2\!\left(\begin{matrix}-n,\xi+iz,\xi-iz\\ \xi+\gamma,\xi+\tau\end{matrix}\bigg|1\right), \quad (3.2.5)$$

with ${}_3F_2\!\left(\begin{matrix}a,b,c\\d,e\end{matrix}\bigg|z\right) = \sum_{n=0}^\infty \frac{(a)_n(b)_n(c)_n}{(d)_n(e)_n}\frac{z^n}{n!}$ and $(c)_n = c(c+1)(c+2)...(c+n-1) = \frac{\Gamma(n+c)}{\Gamma(c)}$. Thus, we conclude that $P_n(\varepsilon) = S_n^\xi(z^2;\gamma,\tau)$ provided that

$$\gamma = \tau = \tfrac{1}{2}(\nu+1),\ \xi = \frac{1}{2} + \frac{2V_1}{\lambda^2}(\varepsilon+M) + \frac{W_1}{\lambda}\left(\frac{2W_0}{\lambda} - 1\right), \quad (3.2.6)$$

and $z^2 = \frac{\varepsilon^2 - \tilde{M}^2}{\lambda^2}$, which requires that $\varepsilon^2 \geq \tilde{M}^2$. The phase shift is obtained using the asymptotics formula of the continuous dual Hahn polynomial. It reads $\delta^\mu(z) = \arg\left[\Gamma(2iz)/\Gamma(\xi+iz)\Gamma(\gamma+iz)\Gamma(\tau+iz)\right]$ giving

$$\delta(\varepsilon) = \arg\Gamma\!\left[2i\lambda^{-1}\sqrt{\varepsilon^2 - \tilde{M}^2}\right] - 2\arg\Gamma\!\left[\frac{\nu+1}{2} + i\lambda^{-1}\sqrt{\varepsilon^2 - \tilde{M}^2}\right]$$
$$- \arg\Gamma\!\left[\frac{1}{2} + \frac{2V_1}{\lambda^2}(\varepsilon+M) + \frac{W_1}{\lambda}\left(\frac{2W_0}{\lambda} - 1\right) + i\lambda^{-1}\sqrt{\varepsilon^2 - \tilde{M}^2}\right] \quad (3.2.7)$$

Now, the spectrum formula of the continuous dual Hahn polynomial reads $z_n^2 = -(n+\xi)^2$, where $n = 0,1,2,..,N$ and $N$ is the largest integer less than or equal to $-\xi$. Consequently, we obtain the following relativistic energy spectrum formula

$$\varepsilon_n^2 = \tilde{M}^2 - \lambda^2\left[ n + \frac{1}{2} + \frac{2V_1}{\lambda^2}(\varepsilon_n + M) + \frac{W_1}{\lambda}\left(\frac{2W_0}{\lambda} - 1\right)\right]^2, \quad (3.2.8)$$

which is a quadratic equation to be solved for $\varepsilon_n$. With $W_0 = 0$, it is identical to the energy spectrum of the spin-symmetric Dirac-Morse problem (i.e., with equal scalar and vector exponential potentials). Now, the continuous dual Hahn polynomial has a mix of continuous and discrete spectra for $\xi < 0$, then the following wavefunction represents the system with a mix of continuous energy $\varepsilon$ and discrete energy $\varepsilon_m$

$$\psi_m^\pm(x,\varepsilon) = f_0(\varepsilon)\sum_n S_n^\xi(z^2;\gamma,\tau)\phi_n^\pm(x) + f_0(\varepsilon_m)\sum_n S_n^\xi(z_m^2;\gamma,\tau)\phi_n^\pm(x), \quad (3.2.9)$$

The corresponding orthogonality relation, which is valid for this case, reads as follows

$$\int_0^\infty \rho^\xi(z;\gamma,\tau) S_n^\xi(z^2;\gamma,\tau) S_{n'}^\mu(z^2;\gamma,\tau) dz + 2 \frac{\Gamma(\gamma-\xi)\Gamma(\tau-\xi)}{\Gamma(\gamma+\tau)\Gamma(1-2\xi)} \times$$
$$\sum_{m=0}^N (-\xi-m) \frac{(\xi+\gamma)_m (\xi+\tau)_m (1-2\xi-m)_m}{(\gamma-\xi-m)_m (\tau-\xi-m)_m m!} S_n^\xi\left(-(m+\xi)^2;\gamma,\tau\right) S_{n'}^\xi\left(-(m+\xi)^2;\gamma,\tau\right) = \delta_{n,n'} \quad (3.2.10)$$

For the second configuration, $y(x) = (\lambda x/2)^2$ and we can write

$$V(r) = V_0 + \frac{4V_1}{\lambda^2} \frac{1}{r^2}, \text{ and } W(r) = \frac{\kappa}{r} + \frac{1}{2}\omega^2 r, \quad (3.2.11)$$

where we have replaced $x$ by the radial coordinate $r$ and $\kappa$ is the spin-orbit coupling constant. To avoid quantum anomalies in the inverse square potential, we demand that its coupling strength be larger than the critical value of $-\frac{1}{4}$ (i.e., $8V_1/\lambda^2 > -\frac{1}{4}$). Additionally, $\omega$ is the pseudo-scaler oscillator frequency and without loss of generality we can always choose $V_0 = 0$. Therefore, the parameters in the Dirac wave operator matrix (A7b) are as follows:

$$\rho = (\omega/\lambda)^2, \ \sigma = \tfrac{1}{2}\kappa, \ A = V_1, \ B = 0 \text{ and } \omega^2 = \tfrac{1}{2}\lambda^2. \quad (3.2.12)$$

Substituting these parameters in (A7b), results in the following symmetric three-term recursion relation for the expansion coefficients of the wavefunction

$$-\left[\frac{2V_1}{\lambda^2}(\varepsilon+M) + \frac{1}{4}\left(\kappa+\tfrac{1}{2}\right)^2\right] f_n = \left\{(2n+\nu+1)\left[n+\frac{\nu}{2}+1+\frac{1}{2}\left(\kappa-\tfrac{1}{2}\right) - \frac{\varepsilon^2-M^2}{\lambda^2}\right] - n - \frac{1}{4}(\nu+1)^2\right\} f_n$$
$$-\left[n+\frac{\nu}{2}+1+\frac{1}{2}\left(\kappa-\tfrac{1}{2}\right) - \frac{\varepsilon^2-M^2}{\lambda^2}\right]\sqrt{(n+1)(n+\nu+1)} f_{n+1} - \left[n+\frac{\nu}{2}+\frac{1}{2}\left(\kappa-\tfrac{1}{2}\right) - \frac{\varepsilon^2-M^2}{\lambda^2}\right]\sqrt{n(n+\nu)} f_{n-1} \quad (3.2.13)$$

Rewriting $f_n(\varepsilon) = f_0(\varepsilon) P_n(\varepsilon)$, makes this a recursion relation for $P_n(\varepsilon)$ with $P_0(\varepsilon) = 1$. Comparing to that of the continuous dual Hahn polynomial (3.2.4) makes $P_n(\varepsilon) = S_n^\xi(z^2;\gamma,\tau)$ such that

$$\gamma = \tau = \tfrac{1}{2}(\nu+1), \ \xi = \frac{1}{2} + \frac{1}{2}\left(\kappa - \tfrac{1}{2}\right) - \frac{\varepsilon^2 - M^2}{\lambda^2}, \quad (3.2.14)$$

and $z^2 = -\left[\frac{2V_1}{\lambda^2}(\varepsilon+M) + \frac{1}{4}\left(\kappa+\tfrac{1}{2}\right)^2\right]$. However, this implies that $z$ is pure imaginary and the system has only discrete bound states. The spectrum formula of the of the continuous dual Hahn polynomial, $z_n^2 = -(n+\xi)^2$, results in the following relativistic energy spectrum

$$\frac{2V_1}{\lambda^2}(\varepsilon_n + M) + \frac{1}{4}\left(\kappa+\tfrac{1}{2}\right)^2 = \left[n + \frac{1}{2} + \frac{1}{2}\left(\kappa-\tfrac{1}{2}\right) - \frac{\varepsilon_n^2 - M^2}{\lambda^2}\right]^2, \quad (3.2.15)$$

which is to be solved for $\varepsilon_n$. With $V_1 = 0$, this is identical to the energy spectrum of the Dirac-oscillator problem for an oscillator frequency $\omega^2 = \tfrac{1}{2}\lambda^2$ and with $\nu = \pm\left(\kappa + \tfrac{1}{2}\right)$ for $\pm\kappa > 0$. The corresponding $m^{th}$ bound state wavefunction will be written in terms of the dual Hahn polynomial $R_n^N(m;\gamma,\tau)$, which is discrete version of the continuous dual Hahn polynomial and defined as

$$R_n^N(m;\gamma,\tau) = \sqrt{\frac{(\gamma+1)_n (\tau+1)_{N-n}}{n!(N-n)!}} \, _3F_2\left(\begin{matrix}-n,-m,m+\gamma+\tau+1\\ \gamma+1,-N\end{matrix}\bigg|1\right), \quad (3.2.16)$$

where $n, m = 0, 1, 2, .., N$ and either $\gamma, \tau > -1$ or $\gamma, \tau < -N$. Therefore, the $m^{th}$ bound state wavefunction is written as $\psi_m^\pm(r) = f_0(\varepsilon_m) \sum_{n=0}^N R_n^N\left(m; \tfrac{\nu+1}{2}, \tfrac{\nu+1}{2}\right) \phi_n^\pm(r)$.

## 4. Solution in the Jacobi basis

Let $y(x)$ be a coordinate transformation such that $-1 \leq y \leq +1$. A complete set of square integrable functions as basis in the new configuration space with the dimensionless coordinate $y$ has the following elements

$$\phi_n(x) = A_n (1-y)^\alpha (1+y)^\beta P_n^{(\mu,\nu)}(y), \tag{4.1}$$

where $P_n^{(\mu,\nu)}(y)$ is the Jacobi polynomial of degree $n$ in $y$ and the normalization constant is chosen as $A_n = \sqrt{\frac{2n+\mu+\nu+1}{2^{\mu+\nu+1}} \frac{\Gamma(n+1)\Gamma(n+\mu+\nu+1)}{\Gamma(n+\mu+1)\Gamma(n+\nu+1)}}$. The real dimensionless parameters $\{\mu,\nu\}$ are greater than $-1$ whereas $\{\alpha,\beta\}$ will be determined by square integrability and the tridiagonal requirement. Substituting (4.1) into Eq. (2.4) and using the differentiation chain rule $\frac{d}{dx} = y' \frac{d}{dy}$, we obtain

$$\phi_n^-(x) = \frac{A_n}{\varepsilon + M} y'(1-y)^{\alpha-1}(1+y)^{\beta-1} \times$$
$$\left[ (1-y^2)\frac{d}{dy} - y(\alpha+\beta) - (\alpha-\beta) + \frac{1-y^2}{y'}W(y) \right] P_n^{(\mu,\nu)}(y) \tag{4.2}$$

where the prime stands for the derivative with respect to $x$. It is required that in function space $\phi_n^-(x)$ must be nearest neighbor to $\phi_n^+(x)$. The differential property of the Jacobi polynomial,

$$(1-y^2)\frac{d}{dy}P_n^{(\mu,\nu)} = -n\left( y + \frac{\nu-\mu}{2n+\mu+\nu} \right)P_n^{(\mu,\nu)} + 2\frac{(n+\mu)(n+\nu)}{2n+\mu+\nu} P_{n-1}^{(\mu,\nu)},$$

and its recursion relation,

$$yP_n^{(\mu,\nu)} = \frac{\nu^2-\mu^2}{(2n+\mu+\nu)(2n+\mu+\nu+2)}P_n^{(\mu,\nu)} + \frac{2(n+\mu)(n+\nu)}{(2n+\mu+\nu)(2n+\mu+\nu+1)}P_{n-1}^{(\mu,\nu)} + \frac{2(n+1)(n+\mu+\nu+1)}{(2n+\mu+\nu+1)(2n+\mu+\nu+2)}P_{n+1}^{(\mu,\nu)},$$

show that this could be achieved if we impose the constraint that $\frac{1-y^2}{y'}W(y)$ be a linear function in $y$. That is, $\frac{1-y^2}{y'}W(y) = \rho(1+y) - \sigma(1-y)$ where $\rho$ and $\sigma$ are dimensionless potential parameters. Thus, we can rewrite Eq. (4.2) as follows

$$\phi_n^-(x) = \frac{y'(1-y^2)^{-1}}{\varepsilon + M} \left\{ \left[ (n+\alpha+\beta-\rho-\sigma)C_n + \frac{n(\mu-\nu)}{2n+\mu+\nu} + (\beta-\alpha+\rho-\sigma) \right] \phi_n^+(x) \right.$$
$$\left. + (n+\mu+\nu-\alpha-\beta+\rho+\sigma+1)D_{n-1}\phi_{n-1}^+(x) - (n+\alpha+\beta-\rho-\sigma)D_n\phi_{n+1}^+(x) \right\} \tag{4.3}$$

where $C_n = \frac{\mu^2-\nu^2}{(2n+\mu+\nu)(2n+\mu+\nu+2)}$ and $D_n = \frac{2}{2n+\mu+\nu+2}\sqrt{\frac{(n+1)(n+\mu+1)(n+\nu+1)(n+\mu+\nu+1)}{(2n+\mu+\nu+1)(2n+\mu+\nu+3)}}$. In the following two subsections, we consider the two physical scenarios found in Appendix B and correspond to equations (B8a) and (B8b), respectively.

### 4.1 The $(p,q) = (0,0)$ scenario of (B8a):

In this scenario, the vector potential takes the form $V(y) = (1-y)^{2a-1}(1+y)^{2b-1}(V_0 + V_1 y)$, and the pseudo-scalar potential reads $W(y) = (1-y)^a (1+y)^b \left[ \frac{W_-}{1-y} - \frac{W_+}{1+y} \right]$, where $\{V_0, V_1, W_\pm\}$ are real potential parameters. Moreover, the basis parameters are restricted to satisfy $\mu^2 = \left( \frac{2W_-}{\lambda} + a - 1 \right)^2$, $\nu^2 = \left( \frac{2W_+}{\lambda} + b - 1 \right)^2$, $2\alpha = \mu + 1 - a$, and $2\beta = \nu + 1 - b$, where $a$ and $b$ are either

–11–

$(a,b) = \left(\frac{1}{2},\frac{1}{2}\right)$ or $(a,b) = \left(0,\frac{1}{2}\right)$. Additionally, the pseudo-scalar potential parameters become $\rho = W_-/\lambda$ and $\sigma = W_+/\lambda$.

For the first case, the solution of $y' = \lambda\sqrt{1-y^2}$ gives $y(x) = \sin(\lambda x)$ with $-\frac{\pi}{2\lambda} < x < \frac{\pi}{2\lambda}$. The basis parameters become $\mu^2 = \left(\frac{2W_-}{\lambda} - \frac{1}{2}\right)^2$, $\nu^2 = \left(\frac{2W_+}{\lambda} - \frac{1}{2}\right)^2$, $2\alpha = \mu + \frac{1}{2}$ and $2\beta = \nu + \frac{1}{2}$. The potential functions read as follows

$$V(x) = V_0 + V_1 \sin(\lambda x) \text{ and } W(x) = (W_+ + W_-)\tan(\lambda x) - \frac{W_+ - W_-}{\cos(\lambda x)}, \quad (4.1.1)$$

where we can always choose $V_0 = 0$. The vector and scalar are potential boxes with sinusoidal bottom whereas the pseudo-scalar is a potential box with $1/x$ singularity of strength $\pm 2W_\mp$ at the two edges of the box. This potential configuration was never reported in literature. Its solution here is a demonstration of the unique advantage of the TRA over other methods for enlarging the class of exactly solvable potentials. Substituting these results back in (B8a), we obtain the following three-term recursion relation for the expansion coefficients

$$2V_1 C_n f_n(\varepsilon) = 2V_1 [D_{n-1} f_{n-1}(\varepsilon) + D_n f_{n+1}(\varepsilon)]$$
$$+ \frac{\lambda^2/4}{\varepsilon + M}\left[(2n+\mu+\nu+1)^2 - 2\left(\mu+\nu+\frac{1}{2}\right) - \frac{\varepsilon^2 - M^2}{\lambda^2/4} - \left(\frac{2W_+}{\lambda} + \frac{2W_-}{\lambda} - 1\right)^2\right] f_n(\varepsilon) \quad (4.1.2)$$

where $C_n$ and $D_n$ are defined in Eq. (4.3). Introducing $P_n(\varepsilon) = \frac{A_0}{f_0(\varepsilon)} \frac{f_n(\varepsilon)}{A_n}$ will transform (4.1.2) to the following recursion relation for $P_n(\varepsilon)$

$$C_n P_n(\varepsilon) = \frac{2(n+\mu)(n+\nu)}{(2n+\mu+\nu)(2n+\mu+\nu+1)} P_{n-1}(\varepsilon) + \frac{2(n+1)(n+\mu+\nu+1)}{(2n+\mu+\nu+1)(2n+\mu+\nu+2)} P_{n+1}(\varepsilon)$$
$$+ \frac{\lambda^2/8V_1}{\varepsilon + M}\left[(2n+\mu+\nu+1)^2 - 2\left(\mu+\nu+\frac{1}{2}\right) - \frac{\varepsilon^2 - M^2}{\lambda^2/4} - \left(\frac{2W_+}{\lambda} + \frac{2W_-}{\lambda} - 1\right)^2\right] P_n(\varepsilon), \quad (4.1.3)$$

Comparing (4.1.3) with Eq. (8) in [9], we conclude that $P_n(\varepsilon) = \bar{H}_n^{(\mu,\nu)}(z^{-1};\alpha,\theta)$, where $\alpha = -\frac{1}{2}\left(\mu+\nu+\frac{1}{2}\right) - \left(\frac{W_+}{\lambda} + \frac{W_-}{\lambda} - \frac{1}{2}\right)^2$, $\cos\theta = \frac{\varepsilon - M}{2V_1}$, and $z = \frac{2V_1}{\lambda^2}(\varepsilon + M)\sin\theta$. Some of the properties of this new polynomial $\bar{H}_n^{(\mu,\nu)}(z^{-1};\alpha,\theta)$ were derived numerically in [9]. In contrast to the orthogonal polynomials of section 3, the analytical properties of this new polynomial are not yet known. Thus, the properties of the corresponding physical system (such as the phase shift and energy spectrum that would have been determined from the asymptotics of the polynomial [9]) could not be given analytically or in a closed form. In the absence of these analytic properties, we give in Table 1 numerical results for the lowest part of the positive energy relativistic spectrum for a chosen set of values of the physical parameters. In Appendix C, we give the details of the procedure used in this calculation. The upper component of the spinor wavefunction is written as $\psi^+(x) = \sum_n f_n(\varepsilon)\phi_n^+(x)$, where $f_n(\varepsilon) = \frac{f_0(\varepsilon)}{A_0} A_n \bar{H}_n^{(\mu,\nu)}(z^{-1};\alpha,\theta)$. The lower component of the spinor wavefunction can be easily obtained by calculating $\phi_n^-(x)$ using Eq. (4.3) with $\rho = W_-/\lambda$ and $\sigma = W_+/\lambda$.

For the second case where $(a,b) = \left(0, \frac{1}{2}\right)$, the solution of $y' = \lambda\sqrt{1+y}$ gives $y(x) = 2(x/L)^2 - 1$ with $0 \leq x \leq L$ and $\lambda = 2\sqrt{2}/L$. The basis parameters become $\mu^2 = (LW_- - 1)^2$, $\nu^2 = \left(LW_+ - \frac{1}{2}\right)^2$, $2\alpha = \mu + 1$ and $2\beta = \nu + \frac{1}{2}$, where we made the replacement $W_\pm \to \sqrt{2}W_\pm$. The solvable potential configuration reads

$$V(x) = \frac{\frac{1}{2}V_0 + V_1\left[(x/L)^2 - \frac{1}{2}\right]}{1-(x/L)^2}, \quad W(x) = \frac{xW_-/L}{1-(x/L)^2} - \frac{LW_+}{x}, \quad (4.1.4)$$

which are potential boxes with $1/x$ singularity at the edges of the box. Using Eq. (B8a), we write the three-term recursion relation associated with this relativistic system as follows

$$C_n f_n = \frac{2/L^2}{(\varepsilon+M)(\varepsilon - M + 2V_1)} \times$$

$$\left[-\frac{L^2}{2}(\varepsilon+M)(\varepsilon-M-2V_0) + (2n+\mu+\nu+1)^2 - 2\left(\mu+\nu+\frac{1}{2}\right) - \left(LW_+ + LW_- - \frac{3}{2}\right)^2\right] f_n \quad (4.1.5)$$

$$+ [D_{n-1} f_{n-1} + D_n f_{n+1}]$$

Following the same procedure as in the previous example, we define $Q_n(\varepsilon) = \frac{A_0}{f_0(\varepsilon)} \frac{f_n(\varepsilon)}{A_n}$, which transforms (4.1.5) to the following form

$$C_n Q_n(\varepsilon) = \frac{2/L^2}{(\varepsilon+M)(\varepsilon-M+2V_1)} \times$$

$$\left[-\frac{L^2}{2}(\varepsilon+M)(\varepsilon-M-2V_0) + (2n+\mu+\nu+1)^2 - 2\left(\mu+\nu+\frac{1}{2}\right) - \left(LW_+ + LW_- - \frac{3}{2}\right)^2\right] Q_n \quad (4.1.6)$$

$$+ \frac{2(n+\mu)(n+\nu)}{(2n+\mu+\nu)(2n+\mu+\nu+1)} Q_{n-1}(\varepsilon) + \frac{2(n+1)(n+\mu+\nu+1)}{(2n+\mu+\nu+1)(2n+\mu+\nu+2)} Q_{n+1}(\varepsilon)$$

Thus, we conclude that $Q_n(\varepsilon) = \bar{H}_n^{(\mu,\nu)}(z^{-1}; \alpha, \theta)$ where $z = \frac{L^2}{8}\sin\theta(\varepsilon+M)(\varepsilon-M+2V_1)$, $\alpha = -\frac{1}{2}\left(\mu+\nu+\frac{1}{2}\right) - \frac{1}{4}\left(LW_+ + LW_- - \frac{3}{2}\right)^2$ and $\cos\theta = \frac{\varepsilon - M - 2V_0}{\varepsilon - M + 2V_1}$ (see Ref. [9] for details).

Again, in the absence of analytic properties of the orthogonal polynomials $Q_n(\varepsilon)$, we give in Table 2 numerical results for the lowest part of the positive energy relativistic spectrum for a chosen set of values of the physical parameters. The spinor wavefunction components can be easily constructed using the same procedure followed in the previous problem.

## 4.2 The $(p,q) = (1,0)$ scenario of (B8b):

In this scenario, the vector potential is $V(y) = (1-y)^{2a-1}(1+y)^{2b-1}\left(\frac{V_-}{1-y} + V_0\right)$ and the pseudo-scalar potential is $W(y) = (1-y)^a(1+y)^b\left(\frac{W_-}{1-y} - \frac{W_+}{1+y}\right)$ where $\rho = W_-/\lambda$ and $\sigma = W_+/\lambda$. Moreover, the basis parameters are $\nu^2 = \left(\frac{2W_+}{\lambda} + b - 1\right)^2$, $2\alpha = \mu + 2 - a$ and $2\beta = \nu + 1 - b$. The parameter $\mu$ is fixed later by physical constraints including the (finite) number of bound states. There are three physical configurations associated with this scenario. The first one corresponds to $(a,b) = \left(1, \frac{1}{2}\right)$ where $y(x) = 2\tanh^2(\lambda x/\sqrt{2}) - 1$ and $x \geq 0$. The second corresponds to

–13–

$(a,b) = \left(\frac{1}{2}, \frac{1}{2}\right)$ where $y(x) = \sin(\lambda x)$ and $-\frac{\pi}{2\lambda} < x < \frac{\pi}{2\lambda}$. The third corresponds to $(a,b) = (1,0)$ where $y(x) = 1 - 2e^{-\lambda x}$ and $x \geq 0$.

For the first case and with $\lambda \to \sqrt{2}\lambda$, the solvable potential configuration reads

$$V(x) = V_- + \frac{2V_0}{\cosh^2(\lambda x)} \text{ and } W(x) = W_- \tanh(\lambda x) - \frac{2W_+}{\sinh(2\lambda x)}, \quad (4.2.1)$$

where we have also made the replacement $W_\pm \to W_\pm/\sqrt{2}$. To force the vector potential to vanish at infinity, we choose $V_- = 0$. The basis parameters become $v^2 = \left(\frac{W_+}{\lambda} - \frac{1}{2}\right)^2$, $2\alpha = \mu + 1$ and $2\beta = v + \frac{1}{2}$. Substituting these quantities in Eq. (B8b) and after somewhat lengthy manipulations, we obtain the following three-term recursion relation

$$\frac{\varepsilon^2 - \hat{M}^2}{2\lambda^2} K_n(\varepsilon) = \left[(\sigma + B_n^2)(C_n + 1) - \frac{2n(n+v)}{2n+\mu+v} - \frac{1}{2}(\mu+1)^2\right] K_n(\varepsilon)$$
$$- (\sigma + B_{n-1}^2) \frac{2(n+\mu)(n+v)}{(2n+\mu+v)(2n+\mu+v+1)} K_{n-1}(\varepsilon) - (\sigma + B_n^2) \frac{2(n+1)(n+\mu+v+1)}{(2n+\mu+v+1)(2n+\mu+v+2)} K_{n+1}(\varepsilon) \quad (4.2.2)$$

where $K_n(\varepsilon) = \frac{A_0}{f_0(\varepsilon)} \frac{f_n(\varepsilon)}{A_n}$, $\hat{M}^2 = M^2 + W_-^2$, $B_n = n + \frac{\mu+v}{2} + 1$ and $\sigma = -\frac{1}{4}\left(\frac{W_+}{\lambda} + \frac{W_-}{\lambda} - \frac{1}{2}\right)^2 + \frac{V_0}{\lambda^2}(\varepsilon + M) - \frac{1}{2}\left(\mu + v + \frac{3}{2}\right)$. Comparing this recursion relation with Eq. (10) in Ref. [9], gives $K_n(\varepsilon) = \bar{G}_n^{(\mu,v)}(z^2; \sigma)$, where $\bar{G}_n^{(\mu,v)}(z^2; \sigma)$ is a new orthogonal polynomial defined in [9] with $2\lambda^2 z^2 = \varepsilon^2 - \hat{M}^2$. Some of the interesting properties of this polynomial are discussed in the same reference [9]. For example, if $\sigma$ is positive then this polynomial has only a continuous spectrum. However, if $\sigma$ is negative then the spectrum is a mix of continuous scattering states and a finite number of discrete bound states. Moreover, the corresponding bound state energies are obtained from the following spectrum formula of the polynomial

$$z_n^2 = -2\left(n + \frac{v+1}{2} - \sqrt{-\sigma}\right)^2, \quad (4.2.3)$$

where $n = 0, 1, ..., N-1$ and $N$ is the largest integer less than or equal to $\sqrt{-\sigma} - \frac{v+1}{2}$.

For the second configuration, $(a,b) = \left(\frac{1}{2}, \frac{1}{2}\right)$, which is equivalent to $y(x) = \sin(\lambda x)$ with $-\frac{\pi}{2\lambda} < x < \frac{\pi}{2\lambda}$, and the potential functions read

$$V(x) = \frac{V_-}{1 - \sin \lambda x} + V_0, \text{ and } W(x) = (W_+ + W_-) \tan(\lambda x) - \frac{W_+ - W_-}{\cos(\lambda x)}, \quad (4.2.4)$$

where $V_- > 0$. It is interesting to note the difference between this case and the potential box in the first case of subsection 4.1 above. The basis parameters become $v^2 = \left(\frac{2W_+}{\lambda} - \frac{1}{2}\right)^2$, $2\alpha = \mu + \frac{3}{2}$ and $2\beta = v + \frac{1}{2}$. Substitution in (B8b), we obtain a three-term recursion relation for this problem that resembles (4.2.2) above and reads as follows

$$z^2 T_n(\varepsilon) = \left[ (\sigma + B_n^2)(C_n + 1) - \frac{2n(n+\nu)}{2n+\mu+\nu} - \tfrac{1}{2}(\mu+1)^2 \right] T_n(\varepsilon) \qquad (4.2.5)$$
$$-(\sigma + B_{n-1}^2) \frac{2(n+\mu)(n+\nu)}{(2n+\mu+\nu)(2n+\mu+\nu+1)} T_{n-1}(\varepsilon) - (\sigma + B_n^2) \frac{2(n+1)(n+\mu+\nu+1)}{(2n+\mu+\nu+1)(2n+\mu+\nu+2)} T_{n+1}(\varepsilon)$$

where $T_n(\varepsilon) = \frac{A_0}{f_0(\varepsilon)} \frac{f_n(\varepsilon)}{A_n}$ and again $B_n = n + \frac{\mu+\nu}{2} + 1$. However, here $z^2 = -\frac{2V_-}{\lambda^2}(\varepsilon + M) - \frac{1}{2}\left(\frac{2W_-}{\lambda} - \frac{1}{2}\right)^2$ and $\sigma = \frac{2V_0}{\lambda^2}(\varepsilon + M) - \frac{\varepsilon^2 - M^2}{\lambda^2} - \frac{1}{4}\left(\frac{2W_+}{\lambda} + \frac{2W_-}{\lambda} - 1\right)^2 - \tfrac{1}{2}\left(\mu + \nu + \tfrac{3}{2}\right)$. Thus, we can write $T_n(\varepsilon) = \bar{G}_n^{(\mu,\nu)}(z^2;\sigma)$. It should be noted that $z^2 < 0$ indicating that the problem has only bound states with energies that are obtained from the spectrum formula (4.2.3). However, the spectrum here is infinite since the spectral terminating condition is satisfied for all integers.

For the last situation where $(a,b) = (1,0)$, we obtain $y(x) = 1 - 2e^{-\lambda x}$ as solution of $y' = \lambda(1-y)$ that satisfy $y \in [-1,+1]$, where $x \geq 0$. The basis parameters become $\nu^2 = \left(\frac{2W_+}{\lambda} - 1\right)^2$, $2\alpha = \mu + 1$ and $2\beta = \nu + 1$. The solvable potential coupling now reads

$$V(x) = \frac{1}{e^{\lambda x} - 1}\left(V_0 + \tfrac{1}{2} V_- e^{\lambda x}\right) \text{ and } W(x) = W_- - \frac{W_+}{e^{\lambda x} - 1}. \qquad (4.2.6)$$

The above potentials are in the form of a generalized Hulthén potential with $x$ being replaced by the radial coordinate $r$. To force the vector potential to vanish at infinity, we must choose $V_- = 0$. Substitution in Eq. (B8b) leads to the following three-term recursion relation for the expansion coefficients of the spinor wavefunction

$$\frac{2}{\lambda^2}(\varepsilon^2 - \hat{M}^2) R_n(\varepsilon) = \left[ (\sigma + B_n^2)(C_n + 1) - \frac{2n(n+\nu)}{2n+\mu+\nu} - \tfrac{1}{2}(\mu+1)^2 \right] R_n(\varepsilon) \qquad (4.2.7)$$
$$-(\sigma + B_{n-1}^2) \frac{2(n+\mu)(n+\nu)}{(2n+\mu+\nu)(2n+\mu+\nu+1)} R_{n-1}(\varepsilon) - (\sigma + B_n^2) \frac{2(n+1)(n+\mu+\nu+1)}{(2n+\mu+\nu+1)(2n+\mu+\nu+2)} R_{n+1}(\varepsilon)$$

where, again, $R_n(\varepsilon) = \frac{A_0}{f_0(\varepsilon)} \frac{f_n(\varepsilon)}{A_n}$, $\hat{M}^2 = M^2 + W_-^2$ and $B_n = n + \frac{\mu+\nu}{2} + 1$. However, $\sigma = \frac{\varepsilon^2 - M^2}{\lambda^2} + \frac{2V_0}{\lambda^2}(\varepsilon + M) - \frac{1}{4}\left(\frac{2W_+}{\lambda} + \frac{2W_-}{\lambda} - 1\right)^2 - \tfrac{1}{2}\left(\mu + \nu + \tfrac{3}{2}\right)$ which could be positive or negative depending on the sign of $V_0$. Therefore, $R_n(\varepsilon) = \bar{G}_n^{(\mu,\nu)}(z^2;\sigma)$ with $\lambda^2 z^2 = 2(\varepsilon^2 - \hat{M}^2)$ and for negative $\sigma$ the bound state energy spectrum is obtained from the spectrum formula (4.2.3). On the other hand, for positive $\sigma$ the system has only continuum scattering states with the two-component wavefunction $\psi^\pm(x) = [f_0(\varepsilon)/A_0] \sum_n A_n R_n(\varepsilon) \phi_n^\pm(x)$ and where the scattering phase shift is obtained from the asymptotics of the polynomial $R_n(\varepsilon)$, or equivalently $\bar{G}_n^{(\mu,\nu)}(z^2;\sigma)$, which is unfortunately not yet know analytically. Consequently, one needs to resort to numerical means.

## 5. Conclusion

In this article, we have discussed different exactly solvable potentials for the Dirac equation that have never been reported in the literature. However, we did not exhaust all possible solvable potentials in this manuscript. For example, we could have included a larger class of potentials by keeping $V_\pm \neq 0$ in the potential $V(x)$ of subsection 4.1 provided that the basis parameters become energy dependent and chosen such that

$$\mu^2 = \left(\tfrac{2W_-}{\lambda}+a-1\right)^2 + \tfrac{4V_-}{\lambda^2}(\varepsilon+M) \text{ and } \nu^2 = \left(\tfrac{2W_+}{\lambda}+b-1\right)^2 + \tfrac{4V_+}{\lambda^2}(\varepsilon+M). \tag{5.1}$$

Additionally, we could have also kept $V_+ \neq 0$ in the potential $V(x)$ of subsection 4.2 provided that the basis parameter $\nu$ is chosen such that

$$\nu^2 = \left(\tfrac{2W_+}{\lambda}+b-1\right)^2 + \tfrac{4V_+}{\lambda^2}(\varepsilon+M). \tag{5.2}$$

Moreover, we did not include the possibility that the basis is neither orthogonal nor tri-thogonal (i.e., the basis overlap matrix $\langle \phi_n^+ | \phi_m^+ \rangle$ is not tridiagonal) but the Dirac wave operator is still tridiagonal. This is accomplished by the requirement that the matrix representation of the kinetic energy operator,

$$\begin{pmatrix} \phi_n^+ & \phi_n^- \end{pmatrix} \begin{pmatrix} M & -\tfrac{d}{dx} \\ \tfrac{d}{dx} & -M \end{pmatrix} \begin{pmatrix} \phi_m^+ \\ \phi_m^- \end{pmatrix}, \tag{5.3}$$

contains a counter term that cancels the non-tridiagonal $\varepsilon\left(\langle \phi_n^+ | \phi_m^+ \rangle + \langle \phi_n^- | \phi_m^- \rangle\right)$.

We also hope that experts in orthogonal polynomials will soon derive the analytical properties of the two orthogonal polynomials mentioned in section 4, which will allow us to write different properties associated with the physical system in closed form, e.g. the energy spectrum and phase shift.

## Acknowledgements:
The authors would like to thank King Fahd University of Petroleum and Minerals (KFUPM) for their support and acknowledge the material support and encouragements of the Saudi Center for Theoretical Physics (SCTP).

## Appendix A: The Laguerre basis

Substituting the Laguerre basis (3.1) into Eq. (2.7) and noting that the integral $\langle f |(...)| g \rangle = \lambda \int_{x_-}^{x_+} f(...)g\, dx = \lambda \int_0^\infty f(...)g\, \tfrac{dy}{y'}$, the last term becomes

$$\left\langle \phi_n^+ \left| \left[ -\tfrac{d^2}{dx^2} + W^2 - \tfrac{dW}{dx} \right] \right| \phi_m^+ \right\rangle =$$

$$-\lambda A_n A_m \left\langle L_n^\nu \left| y^{2\alpha-1} y' e^{-2\beta y} \left[ y \tfrac{d^2}{dy^2} + \left(2\alpha - 2\beta y + \tfrac{yy''}{y'^2}\right)\tfrac{d}{dy} \right. \right. \right. \tag{A1}$$

$$\left. \left. \left. + \tfrac{yy''}{y'^2}\left(\tfrac{\alpha+\sigma}{y}+\rho-\beta\right) - \tfrac{(\sigma+\alpha)(\sigma-\alpha+1)}{y} + y(\beta^2-\rho^2) - 2(\alpha\beta+\rho\sigma) \right] \right| L_m^\nu \right\rangle$$

where $\lambda$ is a length scale parameter to be selected below by an appropriate and natural manner and we have used $W = y'\left(\rho+\tfrac{\sigma}{y}\right)$ and $\tfrac{dW}{dx} = y''\left(\rho+\tfrac{\sigma}{y}\right) - \sigma(y'/y)^2$. Imposing the differential

–16–

equation of the Laguerre polynomial, $\left[ y\frac{d^2}{dy^2} + (v+1-y)\frac{d}{dy} + m \right] L_m^v(y) = 0$, on (A1) dictates that our choice of $y(x)$ should result in $yy''/y'^2$ being linear in $y$. That is, $yy''/y'^2 = a + by$, where $a$ and $b$ are real dimensionless parameters. Writing $z = y'$, this requirement translates into the equation $z'/z^2 = \frac{a}{y} + b$. However, noting that $z' = y'\frac{dz}{dy} = z\frac{dz}{dy}$ we obtain $\frac{dz}{dy}/z = \frac{a}{y} + b$ whose solution is $y' = z = \lambda y^a e^{by}$, where $\lambda$ is the integration constant giving the natural length scale of the problem. Substituting this expression of $y'$ into (A1) gives

$$\left\langle \phi_n^+ \left| -\frac{d^2}{dx^2} + W^2 - \frac{dW}{dx} \right| \phi_m^+ \right\rangle =$$

$$-\lambda^2 A_n A_m \left\langle L_n^v \left| y^{2\alpha+a-1} e^{(b-2\beta)y} \left\{ \left[ 2\alpha + a - v - 1 + y(1+b-2\beta) \right] \frac{d}{dy} \right.\right.\right. \quad \text{(A2)}$$

$$\left. -\frac{(\sigma+\alpha)(\sigma-\alpha-a+1)}{y} + y(\beta-\rho)(\beta+\rho-b) - \left[ m + 2\alpha\beta + 2\rho\sigma + a(\beta-\rho) - b(\sigma+\alpha) \right] \right\} \left| L_m^v \right\rangle$$

Employing the differential property of the Laguerre polynomial, we obtain an off-diagonal matrix element that reads

$$\lambda^2 \sqrt{m(m+v)} \left\langle n \left| y^{2\alpha+a-v-2} e^{(1+b-2\beta)y} \left[ 2\alpha + a - v - 1 + y(1+b-2\beta) \right] \right| m-1 \right\rangle \quad \text{(A3)}$$

where we have defined $\langle y | n \rangle = A_n y^{v/2} e^{-y/2} L_n^v(y) = y^{-\alpha+\frac{1}{2}v} \phi_n^+(x)$. This term must be tridiagonal independent of all other terms or it should vanish. Thus, we must require that $2\beta = 1 + b$ and either (i) $2\alpha = v - a + 2$, or (ii) $2\alpha = v - a + 1$. Combining this with the tridiagonal requirement on the first two terms of Eq. (2.7), we end up with $b = 0$, $\beta = \frac{1}{2}$ and the following two possible scenarios (with $q = 2\alpha + a - v - 1$):

$$q = 0: \quad a = \{0, \tfrac{1}{2}\} \text{ and } V(y) = y^{2a-1}(A + By). \quad \text{(A4a)}$$

$$q = 1: \quad a = \{1, \tfrac{1}{2}\} \text{ and } V(y) = y^{2a-2}(A + By). \quad \text{(A4b)}$$

where $A$ and $B$ are real potential parameters. With these results in (A2), we can write the matrix elements of the Dirac wave operator (2.7) as

$$J_{nm} = 2\langle n | y^{q+1-2a} V(y) | m \rangle - (\varepsilon - M)\langle n | y^{q+1-2a} | m \rangle + \lambda^2 (\varepsilon + M)^{-1} \left\{ q\sqrt{m(m+v)} \langle n | y^{q-1} | m-1 \rangle \right.$$

$$\left. + \langle n | y^q \left[ \frac{(2\sigma+1-a)^2 - (v+q)^2 - 4mq}{4y} + \left(\rho^2 - \tfrac{1}{4}\right)y + \tfrac{1}{2}(2m+v+1+q) + \rho(2\sigma-a) \right] | m \rangle \right\} \quad \text{(A5)}$$

Thus, it is tridiagonal and symmetric in the following two scenarios:

$$q = 0: \quad a = \{0, \tfrac{1}{2}\}, \; v^2 = (2\sigma+1-a)^2 \text{ and } V(y) = y^{2a-1}(A+By). \quad \text{(A6a)}$$

$$q = 1: \quad a = \{1, \tfrac{1}{2}\}, \; \rho^2 = \tfrac{1}{4} \text{ and } V(y) = y^{2a-2}(A+By). \quad \text{(A6b)}$$

For both scenarios the pseudo-scalar potential is $W(y) = \lambda y^{a-1}(\rho y + \sigma)$. Finally, we obtain the following tridiagonal matrix representation of the Dirac wave operator corresponding to the two above scenarios:

$$\lambda^{-2}(\varepsilon + M)J_{n,m} = \left[ \frac{2A}{\lambda^2}(\varepsilon + M) + \tfrac{1}{2}(2m+v+1) + \rho(2\sigma-a) \right]\delta_{n,m}$$

$$+ \left[ \frac{2B}{\lambda^2}(\varepsilon + M) + \left(\rho^2 - \tfrac{1}{4}\right) \right] \langle n | y | m \rangle - \lambda^{-2}(\varepsilon^2 - M^2)\langle n | y^{1-2a} | m \rangle \quad \text{(A7a)}$$

where $a=0$ or $a=\frac{1}{2}$, $v^2=(2\sigma+1-a)^2$ and $2\alpha=v+1-a$.

$$\lambda^{-2}(\varepsilon+M)J_{nm} = \left[\frac{2A}{\lambda^2}(\varepsilon+M)+\left(\sigma+\frac{1-a}{2}\right)^2-\frac{1}{4}(v+1)^2-m\right]\delta_{n,m}$$

$$+\left[\frac{2B}{\lambda^2}(\varepsilon+M)+\frac{1}{2}(2m+v+2)+\rho(2\sigma-a)\right]\langle n|y|m\rangle \qquad (A7b)$$

$$+\sqrt{(n+1)(n+v+1)}\delta_{n,m-1}-\lambda^{-2}(\varepsilon^2-M^2)\langle n|y^{2-2a}|m\rangle$$

where $a=1$ or $a=\frac{1}{2}$, $\rho^2=\frac{1}{4}$ and $2\alpha=v+2-a$. In these expressions, the matrix $\langle n|y|m\rangle$ is obtained using the recursion relation of the Laguerre polynomial and its orthogonality relation, $A_n^2\int_0^\infty y^v e^{-y}L_n^v(y)L_m^v(y)dy=\delta_{nm}$, as follows

$$\langle n|y|m\rangle=(2n+v+1)\delta_{n,m}-\sqrt{n(n+v)}\delta_{n,m+1}-\sqrt{(n+1)(n+v+1)}\delta_{n,m-1}. \qquad (A8)$$

## Appendix B: The Jacobi basis

Substituting the Jacobi basis (4.1) into Eq. (2.7) and noting that the integral $\langle f|(...)|g\rangle = \lambda\int_{x_-}^{x_+}f(...)g\,dx = \lambda\int_0^\infty f(...)g\frac{dy}{y'}$, the last term becomes

$$\left\langle\phi_n^+\left|-\frac{d^2}{dx^2}+W^2-\frac{dW}{dx}\right|\phi_m^+\right\rangle=$$

$$-\lambda A_n A_m\left\langle P_n^{(\mu,v)}\left|(1-y)^{2\alpha-1}(1+y)^{2\beta-1}y'\left\{(1-y^2)\frac{d^2}{dy^2}+\left[2(\beta-\alpha)-2y(\alpha+\beta)+(1-y^2)\frac{y''}{y'^2}\right]\frac{d}{dy}\right.\right.$$

$$+\frac{2\alpha(\alpha-1)}{1-y}+\frac{2\beta(\beta-1)}{1+y}-(\alpha+\beta)(\alpha+\beta-1)-\left[\frac{2\rho(\rho-1)}{1-y}+\frac{2\sigma(\sigma-1)}{1+y}-(\rho+\sigma)(\rho+\sigma-1)\right] \qquad (B1)$$

$$\left.\left.+\frac{y''}{y'^2}[(\beta-\alpha+\rho-\sigma)+y(\rho+\sigma-\alpha-\beta)]\right\}\right|P_m^{(\mu,v)}\right\rangle$$

where we have used $W=y'\left(\frac{\rho}{1-y}-\frac{\sigma}{1+y}\right)$ and $\frac{dW}{dx}=y''\left(\frac{\rho}{1-y}-\frac{\sigma}{1+y}\right)+\frac{y'^2}{1-y^2}\left(\frac{2\rho}{1-y}+\frac{2\sigma}{1+y}-\rho-\sigma\right)$.

Imposing the differential equation of the Jacobi polynomial, $(1-y^2)\frac{d^2}{dy^2}P_n^{(\mu,v)}(y)=[(\mu+v+2)y+\mu-v]\frac{d}{dy}P_n^{(\mu,v)}(y)-n(n+\mu+v+1)P_n^{(\mu,v)}(y)$, on (B1) dictate that $(1-y^2)\frac{y''}{y'^2}$ be linear in $y$. Thus, we may write $(1-y^2)\frac{y''}{y'^2}=-a(1+y)+b(1-y)$ giving $y'=\lambda(1-y)^a\times(1+y)^b$ where $a$ and $b$ are real constants. Substituting this expression of $y'$ into (B1) gives

–18–

$$\left\langle \phi_n^+ \left| -\frac{d^2}{dx^2} + W^2 - \frac{dW}{dx} \right| \phi_m^+ \right\rangle =$$

$$-\lambda^2 A_n A_m \left\langle P_n^{(\mu,\nu)} \left| (1-y)^{2\alpha+a-1}(1+y)^{2\beta+b-1} \left\{ [(2\beta-2\alpha+b-a+\mu-\nu) - y(2\alpha+2\beta+a+b-\mu-\nu-2)] \frac{d}{dy} \right| P_m^{(\mu,\nu)} \right\rangle \right\} \quad \text{(B2)}$$

$$-2\lambda^2 A_n A_m \left\langle P_n^{(\mu,\nu)} \left| (1-y)^{2\alpha+a-1}(1+y)^{2\beta+b-1} \left[ \frac{(\alpha+a+\rho-1)(\alpha-\rho)}{1-y} + \frac{(\beta+b+\sigma-1)(\beta-\sigma)}{1+y} \right] \right| P_m^{(\mu,\nu)} \right\rangle$$

$$+\lambda^2 A_n A_m \left\langle P_n^{(\mu,\nu)} \left| (1-y)^{2\alpha+a-1}(1+y)^{2\beta+b-1} [(\alpha+\beta+a+b+\rho+\sigma-1)(\alpha+\beta-\rho-\sigma) + m(m+\mu+\nu+1)] \right| P_m^{(\mu,\nu)} \right\rangle$$

Employing the differential property of the Jacobi polynomial, we obtain an off-diagonal matrix element that reads

$$\lambda^2 (2m+\mu+\nu+1) D_{m-1} \left\langle n \left| (1-y)^p (1+y)^q \left[ \frac{p}{1-y} - \frac{q}{1+y} \right] \right| m-1 \right\rangle \quad \text{(B3)}$$

where $p = 2\alpha + a - \mu - 1$, $q = 2\beta + b - \nu - 1$, $D_n = \frac{2}{2n+\mu+\nu+2} \sqrt{\frac{(n+1)(n+\mu+1)(n+\nu+1)(n+\mu+\nu+1)}{(2n+\mu+\nu+1)(2n+\mu+\nu+3)}}$ and we have defined $\langle y | n \rangle = A_n (1-y)^{\mu/2}(1+y)^{\nu/2} P_n^{(\mu,\nu)}(y)$. This term must be tridiagonal independent of all other terms or it should vanish. This leads to three possibilities: $(p,q) = (0,0)$ or $(p,q) = (0,1)$ or $(p,q) = (1,0)$. Combining this with the tridiagonal requirement on the first two terms of Eq. (2.7) dictates that $V(y) = (1-y)^{2a-1}(1+y)^{2b-1}$

$\left[ \frac{V_+}{1+y} + \frac{V_-}{1-y} + V_0 + V_1 y \right]$ and we end up with the following three possible scenarios:

$(p,q) = (0,0)$: $(a,b) = \left( \frac{1}{2}, \frac{1}{2} \right)$ or $\left( \frac{1}{2}, 0 \right)$ or $\left( 0, \frac{1}{2} \right)$ and $V_\pm = 0$. (B4a)

$(p,q) = (1,0)$: $(a,b) = \left( 1, \frac{1}{2} \right)$ or $(1,0)$ or $\left( \frac{1}{2}, \frac{1}{2} \right)$ and $V_+ = V_1 = 0$. (B4b)

$(p,q) = (0,1)$: $(a,b) = \left( \frac{1}{2}, 1 \right)$ or $\left( \frac{1}{2}, \frac{1}{2} \right)$ or $(0,1)$ and $V_- = V_1 = 0$. (B4c)

where $\{V_\pm, V_0, V_1\}$ are real potential parameters. For all three scenarios the pseudo-scalar potential is $W(y) = (1-y)^a (1+y)^b \left[ \frac{W_-}{1-y} - \frac{W_+}{1+y} \right]$ where $\lambda \rho = W_-$ and $\lambda \sigma = W_+$. With these results in (B2), we can write the matrix elements of the Dirac wave operator (2.7) as

$$\lambda^{-2}(\varepsilon + M) J_{nm} = 2\lambda^{-2}(\varepsilon + M) \langle n | (1-y)^{p-2a+1}(1+y)^{q-2b+1} V(y) | m \rangle - \lambda^{-2}(\varepsilon^2 - M^2) \langle n | (1-y)^{p-2a+1}(1+y)^{q-2b+1} | m \rangle$$

$$+(2m+\mu+\nu+1) D_{m-1} \left\langle n \left| (1-y)^p (1+y)^q \left[ \frac{p}{1-y} - \frac{q}{1+y} \right] \right| m-1 \right\rangle + \left\langle n \left| (1-y)^p (1+y)^q \left[ \frac{p}{1-y} - \frac{q}{1+y} \right] \left[ \frac{m(\mu-\nu)}{2m+\mu+\nu} - my \right] \right| m \right\rangle$$

$$+ \frac{1}{4} \left\langle n \left| (1-y)^p (1+y)^q \left\{ \frac{2}{1-y} \left[ \left( \frac{2W_-}{\lambda} + a - 1 \right)^2 - (\mu+p)^2 \right] + \frac{2}{1+y} \left[ \left( \frac{2W_+}{\lambda} + b - 1 \right)^2 - (\nu+q)^2 \right] \right. \right. \quad \text{(B5)}$$

$$\left. \left. + (\mu+\nu+p+q)^2 - \left( \frac{2W_+}{\lambda} + \frac{2W_-}{\lambda} + a + b - 2 \right)^2 + (2m+\mu+\nu+1)^2 - (\mu+\nu+1)^2 \right\} \right| m \right\rangle$$

We observe a symmetry in the three scenarios above that allows us to obtain the solution corresponding to an $(a,b)$ case by a simple parameter map from another $(b,a)$ case. One can show that any $(a,b)$ case belonging to (B4c) is obtained from $(b,a)$ case in (B4b) by the following simple map

$y \to -y$, $\lambda \to -\lambda$, $V_\pm \to V_\mp$, $V_1 \to -V_1$, $W_\pm \to -W_\mp$, $\mu \leftrightarrow \nu$, and $\alpha \leftrightarrow \beta$. (B6)

Additionally, any $(a,b)$ case belonging to (B4a) is obtained from $(b,a)$ case in (B4a) by the same map. Thus, we need to deal with only five cases in the following two scenarios:

$(p,q) = (0,0)$: $(a,b) = \left( \frac{1}{2}, \frac{1}{2} \right)$ or $\left( 0, \frac{1}{2} \right)$ and $V_\pm = 0$. (B7a)

$(p,q) = (1,0)$: $(a,b) = \left( 1, \frac{1}{2} \right)$ or $(1,0)$ or $\left( \frac{1}{2}, \frac{1}{2} \right)$ and $V_+ = V_1 = 0$. (B7b)

Finally, we obtain the following tridiagonal matrix representation of the Dirac wave operator corresponding to the two scenarios of (B7) above:

$$\lambda^{-2}(\varepsilon + M)J_{nm} =$$

$$2\lambda^{-2}(\varepsilon + M)\langle n|(V_0 + V_1 y)|m\rangle - \lambda^{-2}(\varepsilon^2 - M^2)\langle n|(1-y)^{1-2a}(1+y)^{1-2b}|m\rangle \quad \text{(B8a)}$$

$$+\frac{1}{4}\left[(2n+\mu+\nu+1)^2 - 2(\mu+\nu+\tfrac{1}{2}) - \left(\tfrac{2W_+}{\lambda} + \tfrac{2W_-}{\lambda} + a + b - 2\right)^2\right]\delta_{nm}$$

where $(a,b) = \{(\tfrac{1}{2},\tfrac{1}{2}),(0,\tfrac{1}{2})\}$, $\mu^2 = \left(\tfrac{2W_-}{\lambda} + a - 1\right)^2$, $\nu^2 = \left(\tfrac{2W_+}{\lambda} + b - 1\right)^2$, $2\alpha = \mu + 1 - a$ and $2\beta = \nu + 1 - b$.

$$\lambda^{-2}(\varepsilon + M)J_{nm} =$$

$$2\lambda^{-2}(\varepsilon + M)\langle n|[V_- + V_0(1-y)]|m\rangle - \lambda^{-2}(\varepsilon^2 - M^2)\langle n|(1-y)^{2-2a}(1+y)^{1-2b}|m\rangle$$

$$+(2m+\mu+\nu+1)D_{m-1}\delta_{n,m-1} - \frac{1}{4}\left[(2m+\mu+\nu+2)^2 - 2(\mu+\nu+1) - 1 - \left(\tfrac{2W_+}{\lambda} + \tfrac{2W_-}{\lambda} + a + b - 2\right)^2\right]\langle n|y|m\rangle \quad \text{(B8b)}$$

$$+\frac{1}{2}\left\{\left(\tfrac{2W_-}{\lambda} + a - 1\right)^2 - (\mu+1)^2 + \frac{1}{2}\left[(2n+\mu+\nu+1)^2 - \left(\tfrac{2W_+}{\lambda} + \tfrac{2W_-}{\lambda} + a + b - 2\right)^2\right] + \frac{2n(\mu-\nu)}{2n+\mu+\nu}\right\}\delta_{n,m}$$

where $(a,b) = \{(1,\tfrac{1}{2}),(1,0),(\tfrac{1}{2},\tfrac{1}{2})\}$, $\nu^2 = \left(\tfrac{2W_+}{\lambda} + b - 1\right)^2$, $2\alpha = \mu + 2 - a$ and $2\beta = \nu + 1 - b$. In these expressions, the matrix $\langle n|y|m\rangle$ is obtained using the recursion relation of the Jacobi polynomial and its orthogonality relation, $A_n^2 \int_{-1}^{+1}(1-y)^\mu(1-y)^\nu P_n^{(\mu,\nu)}(y)P_m^{(\mu,\nu)}(y)dy = \delta_{n,m}$, as follows

$$\langle n|y|m\rangle = -C_n \delta_{n,m} + D_{n-1}\delta_{n,m+1} + D_n \delta_{n,m-1}, \quad \text{(B9)}$$

where the coefficients $C_n$ and $D_n$ are defined below Eq. (4.3).

## Appendix C: Energy spectrum calculation in the Jacobi basis

We choose the lower component of the spinor basis to be energy independent and be related to the upper component via a relation similar to the kinetic balance equation and as follows

$$\phi_n^-(x) = \frac{1}{\tau}\left[\frac{d}{dx} + W(x)\right]\phi_n^+(x), \quad \text{(C1)}$$

where $\tau$ is a non-physical computational parameter of inverse length dimension. We expect that physical results will be independent of the choice of value of this parameter as long as that choice is either unique or natural. Since the choice of $\tau$ in the formulation of the problem above is $\varepsilon + M$ (see below Eq. (2,6)), then we expect that this unique or natural choice of $\tau$ will be different for each energy eigenvalue. If we designate the value of the $m^{\text{th}}$ bound state as a function of $\tau$ as $\varepsilon_m(\tau)$, then the unique or natural choice of $\tau$ is $\bar{\tau}_m$ at which $\tfrac{d}{d\tau}\varepsilon_m(\tau)|_{\tau=\bar{\tau}_m} = 0$. That is, the energy spectrum as a function of $\tau$ is at an extremum. Now, in this energy independent basis the matrix elements of the Dirac wave operator (2.6) become

$$J_{nm} = 2\langle\phi_n^+|V|\phi_m^+\rangle - (\varepsilon - M)\langle\phi_n^+|\phi_m^+\rangle + (2\tau - \varepsilon - M)\langle\phi_n^-|\phi_m^-\rangle. \quad \text{(C2)}$$

Substituting for $\phi_n^-$ using (C1) and following the same procedure that lead to the matrix elements of the wav operator in Appendix B, we obtain

$$J_{nm} = 2\langle n|(1-y)^{p-2a+1}(1+y)^{q-2b+1}V(y)|m\rangle - (\varepsilon - M)\langle n|(1-y)^{p-2a+1}(1+y)^{q-2b+1}|m\rangle$$
$$+ (\lambda/\tau)^2 (2\tau - \varepsilon - M)\mathcal{J}_{nm} \quad \text{(C3a)}$$

Where

$$\mathcal{J}_{nm} = (2m+\mu+\nu+1)D_{m-1}\langle n|(1-y)^p(1+y)^q\left[\frac{p}{1-y}-\frac{q}{1+y}\right]|m-1\rangle + \langle n|(1-y)^p(1+y)^q\left[\frac{p}{1-y}-\frac{q}{1+y}\right]\left[\frac{m(\mu-\nu)}{2m+\mu+\nu}-my\right]|m\rangle$$

$$+\frac{1}{4}\langle n|(1-y)^p(1+y)^q\Big\{\frac{2}{1-y}\left[\left(\frac{2W_-}{\lambda}+a-1\right)^2-(\mu+p)^2\right]+\frac{2}{1+y}\left[\left(\frac{2W_+}{\lambda}+b-1\right)^2-(\nu+q)^2\right] \quad \text{(C3b)}$$

$$+(\mu+\nu+p+q)^2-\left(\frac{2W_+}{\lambda}+\frac{2W_-}{\lambda}+a+b-2\right)^2+(2m+\mu+\nu+1)^2-(\mu+\nu+1)^2\Big\}|m\rangle$$

For the case corresponding to $(p,q) = (0,0)$ the matrix elements of the wave operator becomes

$$J_{nm} = 2\langle n|(V_0+V_1 y)|m\rangle - (\varepsilon - M)\langle n|(1-y)^{1-2a}(1+y)^{1-2b}|m\rangle$$
$$+ (\lambda/\tau)^2(2\tau-\varepsilon-M)\left[\left(n+\frac{\mu+\nu+1}{2}\right)^2 - \frac{1}{2}\left(\mu+\nu+\frac{1}{2}\right) - \left(\frac{W_+}{\lambda}+\frac{W_-}{\lambda}+\frac{a+b}{2}-1\right)^2\right]\delta_{nm} \quad \text{(C4)}$$

On the other hand, the case corresponding to $(p,q) = (1,0)$ the matrix elements of the wave operator becomes

$$J_{nm} = 2\langle n|[V_- + V_0(1-y)]|m\rangle - (\varepsilon-M)\langle n|(1-y)^{2-2a}(1+y)^{1-2b}|m\rangle$$

$$+\left(\frac{\lambda}{\tau}\right)^2(2\tau-\varepsilon-M)\Bigg\{\left[\frac{1}{2}\left(\frac{2W_-}{\lambda}+a-1\right)^2-\frac{1}{2}(\mu+1)^2+\left(n+\frac{\mu+\nu+1}{2}\right)^2-\left(\frac{W_+}{\lambda}+\frac{W_-}{\lambda}+\frac{a+b}{2}-1\right)^2+\frac{n(\mu-\nu)}{2n+\mu+\nu}\right]\delta_{n,m} \quad \text{(C5)}$$

$$+(2m+\mu+\nu+1)D_{m-1}\delta_{n,m-1}+\left[\frac{1}{2}(\mu+\nu+1)+\frac{1}{4}-\left(m+\frac{\mu+\nu}{2}+1\right)^2+\left(\frac{W_+}{\lambda}+\frac{W_-}{\lambda}+\frac{a+b}{2}-1\right)^2\right]\langle n|y|m\rangle\Bigg\}$$

Now, the energy spectrum is calculated using the wave equation $H|\psi\rangle = \varepsilon|\psi\rangle$ as the generalized eigenvalues $\{\varepsilon\}$ of the matrix equation $\sum_m H_{n,m} f_m = \varepsilon \sum_m \Omega_{n,m} f_m$, where the Hamiltonian matrix is obtained from the wave operator matrix (C4) or (C5) as $H = J\big|_{\varepsilon=0}$ and the basis overlap matrix elements, $\Omega_{n,m} = \langle \phi_n|\phi_m\rangle$, are those that are proportional to $-\varepsilon$ in (C4) or (C5).

To test the procedure, we use (C5) to calculate the energy spectra for all three problems of subsection 4.2 and compare them to those obtained using the exact spectrum formula (4.2.3). Agreement is achieved to machine accuracy for large enough basis size. Consequently, we employ the same procedure but using (C4) to obtain the energy spectra for the two problems of subsection 4.1 that do not have an exact spectrum formula. Those results are shown in Table 1 and Table 2.

## Table Caption:

**Table 1**: The lowest part of the energy spectrum associated with the potential configuration (4.1.1) for various basis sizes. We used the procedure outlined in Appendix C and took the following values of the physical parameters: $M=1$, $\lambda=1$, $V_0=0$, $V_1=5$, $W_+=-2$, and $W_-=3$.

**Table 2**: The lowest part of the energy spectrum associated with the potential configuration (4.1.4) for various basis sizes. We took the following values of the physical parameters: $M=1$, $L=1$, $V_0=5$, $V_1=-4$, $W_+=-2$, and $W_-=3$.

### Table 1

| $n$ | $10\times 10$ | $20\times 20$ | $50\times 50$ | $100\times 100$ |
|---|---|---|---|---|
| 0 | 3.4699133741 | 3.4699133741 | 3.4699133741 | 3.4699133741 |
| 1 | 4.9682581470 | 4.9682581410 | 4.9682581410 | 4.9682581410 |
| 2 | 6.3787433799 | 6.3787403137 | 6.3787403137 | 6.3787403137 |
| 3 | 7.7072983711 | 7.7069987810 | 7.7069987810 | 7.7069987810 |
| 4 | 8.9714973010 | 8.9632791265 | 8.9632791265 | 8.9632791265 |
| 5 | 10.2363637462 | 10.1585749545 | 10.1585749545 | 10.1585749545 |
| 6 | 11.6326442412 | 11.3036671639 | 11.3036671639 | 11.3036671639 |
| 7 | 13.2590559338 | 12.4087556852 | 12.4087556851 | 12.4087556851 |
| 8 | 15.1884790045 | 13.4830950092 | 13.4830950080 | 13.4830950080 |
| 9 | 17.6308087613 | 14.5346195400 | 14.5346194735 | 14.5346194735 |

### Table 2

| $n$ | $10\times 10$ | $20\times 20$ | $50\times 50$ | $100\times 100$ |
|---|---|---|---|---|
| 0 | 13.4190303994 | 13.4190303994 | 13.4190303994 | 13.4190303994 |
| 1 | 16.2519177286 | 16.2519177286 | 16.2519177286 | 16.2519177286 |
| 2 | 19.2243051783 | 19.2243051756 | 19.2243051756 | 19.2243051756 |
| 3 | 22.2642542944 | 22.2642446028 | 22.2642446028 | 22.2642446028 |
| 4 | 25.3441540571 | 25.3408807911 | 25.3408807911 | 25.3408807911 |
| 5 | 28.5842009466 | 28.4392304686 | 28.4392304686 | 28.4392304686 |
| 6 | 32.9359626980 | 31.5512768056 | 31.5512768056 | 31.5512768056 |
| 7 | 40.5489349697 | 34.6723962038 | 34.6723962033 | 34.6723962033 |
| 8 | 56.2347940964 | 37.7997605078 | 37.7997601910 | 37.7997601910 |
| 9 | 101.4111438243 | 40.9316226010 | 40.9315548628 | 40.9315548628 |